\documentclass[reprint,nofootinbib,aps,superscriptaddress,amsmath,amssymb,onecolumn,11pt]{revtex4-2}

\usepackage{graphicx}
\usepackage{dcolumn}
\usepackage{bm}
\usepackage{braket}
\usepackage{breqn}
\usepackage{mathrsfs}
\usepackage{color}

\usepackage[colorlinks=true,linkcolor=red,citecolor=blue]{hyperref}

\begin{document}
    
\title{Tachyonic instability and spontaneous scalarization in parameterized Schwarzschild-like black holes}

\author{Hengyu Xu}
\email{xuhengyu0501@outlook.com}
\author{Yizhi Zhan}
\email{niannian$\_$12138@163.com}
\author{Shao-Jun Zhang}
\email{sjzhang@zjut.edu.cn (corresponding author)}
\affiliation{Institute for Theoretical Physics and Cosmology$,$ Zhejiang University of Technology$,$ Hangzhou 310032$,$ China}
\affiliation{United Center for Gravitational Wave Physics$,$ Zhejiang University of Technology$,$ Hangzhou 310032$,$ China}
\date{\today}

\begin{abstract}
    We study the phenomenon of spontaneous scalarization in parameterized Schwarzschild-like black holes. Two metrics are considered, the Konoplya-Zhidenko metric and the Johannsen-Psaltis metric. While these metrics can mimic the Schwarzschild black hole well in the weak-field regime, they have deformed geometries in the near-horizon strong-field region. Such deformations notably influence the emergence of tachyonic instability and subsequent spontaneous scalarization, enabling a clear distinction between these parameterized metrics and the standard Schwarzschild metric. These results suggest a possible way to test the parameterized black holes and thus the Kerr hypothesis by observing the phenomenon of spontaneous scalarization.
\end{abstract}

\maketitle

\section{Introduction}

In recent years, there has been a notable focus on spontaneous scalarization of black holes. This phenomenon often involves a non-minimal coupling between a scalar field and the gravitational field. The scalar field will acquire an effective mass through this coupling on the background of a hairless black hole spacetime. Under certain conditions, the square of this effective mass can become negative near the event horizon, causing a tachyonic instability that results in the evolution of hairless black holes to hairy black holes. This intriguing phenomenon provides a natural mechanism to endow black holes with hair. This phenomenon was first discovered a long time ago in the context of neutron stars \cite{Damour:1993hw}, but there the instability is induced by the surrounding matter rather than the curvature. Subsequently, within the framework of Einstein-scalar-Gauss-Bonnet gravity, it was revealed by researchers that this unique phenomenon also takes place in Schwarzschild black holes \cite{Doneva:2017bvd, Silva:2017uqg, Antoniou:2017acq}. Building upon these seminal studies, researchers have undertaken extensive explorations in this area, extending the work to Kerr black holes \cite{Cunha:2019dwb, Herdeiro:2020wei, Berti:2020kgk}, various other spontaneous scalarization models \cite{Gao:2018acg, Myung:2020etf, Doneva:2021dcc, Zhang:2021btn, Zhang:2022sgt, Herdeiro:2018wub, Zhang:2021nnn} and binary systems \cite{Barausse:2012da, Doneva:2022byd, Elley:2022ept, Palenzuela:2013hsa, Shibata:2013pra, Silva:2020omi, Taniguchi:2014fqa}. For more details, refer to a recent review \cite{Doneva:2022ewd} and the references therein. These studies imply the universality of spontaneous scalarization in modified gravity theories. Notably, black holes adorned with hair exhibit distinct astronomical phenomena compared to their hairless counterparts, impacting phenomena such as black hole shadows and gravitational wave emissions during binary mergers. Theoretical analysis of these observational effects has been conducted \cite{Cunha:2019dwb, Wong:2022wni}, with the potential for future astronomical observations, especially in strong gravity regimes, to validate the existence of these effects and evaluate the physical implications of such a phenomenon.

On the other hand, general relativity (GR) and the Kerr hypotheses still need further testing, especially in strong-field regimes. It is widely believed that, in the near-horizon region, the geometry of a black hole should be modified to some extent to deviate from the Kerr metric. This correction may come from quantum effects or extra dimension scenarios. To model such modifications classically, non-Kerr black hole solutions are developed as an effective way. 

The construction of non-Kerr black holes can be approached through two main methods: the top-down and bottom-up approaches. In the top-down approach, researchers start with modified gravity theories, which have been developed over recent decades to address inconsistencies between GR and cosmological observations like dark matter and dark energy. These modified theories lead to non-Kerr black holes as solutions to the gravitational field equations, but finding exact solutions is often difficult and limited to specific cases or approximations. Examples include Einstein-dilaton-Gauss-Bonnet \cite{Kanti:1995vq, Ayzenberg:2014aka, Maselli:2015tta, Kleihaus:2015aje, Kokkotas:2017ymc}, Chern-Simons \cite{Yunes:2009hc, Yagi:2012ya, McNees:2015srl, Delsate:2018ome}, and Kerr-Sen black holes \cite{Sen:1992ua}. In contrast, the bottom-up approach involves generalizing the Kerr metric directly to obtain non-Kerr black holes without solving complex field equations \cite{Vigeland:2011ji,Johannsen:2011dh,Johannsen:2013szh,Rezzolla:2014mua,Konoplya:2016jvv, Papadopoulos:2018nvd, Carson:2020dez, Konoplya:2023owh, Johannsen:2011dh}. This method resembles the parametrized post-Newtonian (PPN) approach in weak gravity regimes \cite{Will:2014kxa}, and resulting black hole metrics are theory-independent. Some non-Kerr metrics derived from these approaches closely match Kerr black holes in weak-field regimes, suggesting that they are consistent with current astronomical observations, highlighting the importance of considering non-Kerr black holes due to limitations in current observational accuracy.

Investigating the distinguishing Kerr black holes from non-Kerr black holes, both theoretically and experimentally, is a compelling topic in the realm of gravitational physics. The primary discrepancy between the two lies within the strong gravity region proximate to the event horizon, suggesting that the phenomenon of spontaneous scalarization may serve as a viable method of differentiation. Motivated by preceding research, our focus will be on probing the occurrence of spontaneous scalarization in non-Kerr black holes, particularly emphasizing parameterized black hole models. To concisely illustrate the physics involved, we will consider two classes of spherically symmetric Schwarzschild-like black holes, described by the Konoplya-Zhidenko (KZ) metric \cite{Konoplya:2023owh} and the Johannsen-Psaltis (JP) metric \cite{Johannsen:2011dh}, respectively. There are usually two popular ways to construct parametrized black hole metrics. One adopts a continued-fraction expansion in terms of a compactified radial coordinate, while the other is based on a Taylor expansion in terms of $1/r$. The KZ and JP metrics are the typical metrics constructed by the two ways respectively. For more discussions on their differences, one can refer to \cite{Konoplya:2016jvv}. Both agree well with the Schwarzschild metric in the weak-field region and thus can serve as good alternatives to the Schwarzschild metric at the current accuracy of astronomical observations. Moreover, they have many merits, for example, the KZ metric needs only three deformation parameters to fully describe very general Schwarzschild-like black holes, while the JP metric is more concise in form. In this paper, we will see that deviations in the near-horizon geometry will exert a significant influence on the spontaneous scalarization. Future observations of these effects are expected to place more stringent constraints on these deformations, thus testing GR and the Kerr hypothesis.

The work is organized as follows. In Sec. II, we will give a brief introduction to our model where the scalar field is non-minimally coupled to the gravitational field. Then, in Sec. III and Sec. IV, we consider two types of parametrized Schwarzschild-like black holes described by the KZ and JP metrics, respectively, and study the influences of deformation parameters on the occurrence of tachyonic instability and the associated spontaneous scalarization. The last section is the summary and discussions.

\section{the model}

In the parameterized black hole background, we consider a real scalar field non-minimally coupled to the gravity with the action
\begin{equation}
    S = \int dx^4\sqrt{-g}\left[-\frac{1}{2}\partial_\mu\phi\partial^\mu\phi +  \alpha f(\phi) \mathcal{G}\right],
\end{equation}
where $\mathcal{G} = {\cal R}^2+{\cal R}_{\mu\nu\rho\sigma}{\cal R}^{\mu\nu\rho\sigma}-4{\cal R}_{\mu\nu}{\cal R}^{\mu\nu}$ is the Gauss-Bonnet term, $\alpha$ is the coupling constant and $f(\phi)$ is the coupling function. From the action, one can derive the equation of motion for the scalar field,
\begin{equation}
    \nabla^2\phi = -\alpha\frac{d f(\phi)}{d \phi}{\cal G}\label{ScalarEq1}.
\end{equation}
We take the form of the coupling function as \cite{Doneva:2017bvd,Cunha:2019dwb}
\begin{equation}
    f(\phi) = \frac{1}{2\beta}(1 - e^{-\beta \phi^2}),
\end{equation}
where $\beta > 0$ is a constant. Then the scalar field equation (\ref{ScalarEq1}) becomes
\begin{equation}
    \nabla^2\phi = -\alpha {\cal G} \phi\label{ScalarEq2},
\end{equation}
where the Gauss-Bonnet term ${\cal G}$ is valued in the parameterized black hole background. From it one can see that the scalar field acquires an effective mass whose square is $m^2_{\rm eff} =-\alpha {\cal G} $. The effective mass square $m^2_{\rm eff}$ depends on the explicit form of the coupling function $f(\phi)$ as well as the coupling constant $\alpha$ and is position-dependent. When $m^2_{\rm eff} < 0$ somewhere, tachyonic instability may be triggered resulting in the associated spontaneous scalarization. 

We will consider two types of parameterized spherical black holes; one is the Konoplya-Zhidenko metric proposed in \cite{Konoplya:2023owh} and the other is the Johannsen-Psaltis metric proposed in \cite{Johannsen:2011dh}. Both metrics take the form of
\begin{align}
    ds^2 = - g(r) dt^2 + \frac{dr^2}{\chi(r)} + r^2 (d\theta^2 + \sin^2 \theta d\varphi^2). \label{Metric}
\end{align}
The location of the event horizon $r=r_0$ is given by the largest root of $\chi(r)=0$. Several parameters in each type characterize the deformations from the standard Schwarzschild black hole, especially in the near-horizon region. When all deformation parameters vanish, both metrics recover the standard Schwarzschild black hole exactly. We will focus on the influence of these deformations on the tachyonic instability and the associated spontaneous scalarization. As we shall see in the following, different deformations will leave different imprints on this phenomenon, which thus may provide us a way to distinguish different black hole models and test the Kerr hypothesis.

\section{Konoplya-Zhidenko metric}
The Konoplya-Zhidenko (KZ) metric \cite{Konoplya:2023owh} takes the form (\ref{Metric}) with  
\begin{align}
    g(r)&=\left(1-\frac{r_0}{r}\right) \left(1-\frac{r_0 \epsilon}{r}-\frac{r_0^2 \epsilon}{r^2}+\frac{r_0^3}{r^2}\frac{a_1}{r+a_2(r-r_0)}\right), \quad \chi(r) = \frac{g(r)}{B^2(r)},\nonumber\\
    B^2(r)&=\left(1+\frac{r_0^2}{r}\frac{b_1}{r+b_2(r-r_0)}\right)^2,
\end{align}
where $r=r_0$ is the location of the event horizon. This metric describes a general spherically symmetric black hole with five deformation parameters, $\{\epsilon, a_1, a_2, b_1, b_2\}$, which are introduced to characterize deviations from the standard Schwarzschild metric. It reduces to the Schwarzschild metric exactly when all deformation parameters vanish. The horizon deviation parameter $\epsilon$ measures the deviation of $r_0$ from the Schwarzschild radius $r=2 M$,
\begin{equation}
    \epsilon=\frac{2M-r_0}{r_0}.
\end{equation}
The five deformation parameters are not independent of each other. For this metric to serve as a good mimicker of the Schwarzschild metric in the weak-field regime, these deformation parameters should satisfy the following relations
\begin{align}
    a_1 &= -(3+a_2)\epsilon\label{a1},\\
    b_1 &= -\frac{4(2+a_2)(3+b_2)}{(3+a_2)^2}\epsilon\label{a1b1Relations}.
\end{align}
Therefore, there are only three independent deformation parameters $\{\epsilon, a_2, b_2\}$. Current estimations for the 2PN parameters from the binary black hole gravitational signals \cite{LIGOScientific:2019fpa} suggest that
\begin{equation}
    |a_1| \lesssim 1 + a_2,     
    \quad |b_1| \lesssim 1 + b_2. \label{a1b1}
\end{equation}
On the other hand, to make sure that the metric coefficients $g(r)$
and $\chi(r)$ are positive definite outside the event horizon and to avoid naked singularities, the deformation parameters need also satisfy the following constraints
\begin{align}
    &a_2 > -1,
    \quad b_2 > -1\label{a2b2min},\\
    &a_1 >2 \epsilon -1,
    \quad 1 + b_1 > 0.
    \label{constraints}
\end{align}
With the relations (\ref{a1b1Relations}), the above constraints give us
\begin{equation}
\left\{\begin{array}{cc}
    a_2 >\frac{1}{\epsilon}-5, 
   & b_2 >\frac{(3+a_2)^2}{4(2+a_2)\epsilon}-3, \quad {\rm when } \ \epsilon < 0 ,\\
    a_2 < \frac{1}{\epsilon}-5,
   & b_2 <\frac{(3+a_2)^2}{4(2+a_2)\epsilon}-3, \quad {\rm when }\ \epsilon > 0.
\end{array}\right.
\end{equation}

To mimic the shadow of the Schwarzschild black hole with sufficient accuracy, $\epsilon$ has to be small \cite{Konoplya:2023owh}. In the following, we assume a reasonable range of values for $\epsilon$, $\epsilon \in [-0.04, 0.04]$. 

Let us first do a primary analysis on the onset of the tachyonic instability. Considering that the background is static and spherically symmetric, the scalar field perturbation can be decomposed as
\begin{equation}
    \phi = \frac{\psi(r)}{r} Y_{\ell m} (\theta,\varphi) e^{- i\omega t},\label{ScalarDecomposition}
\end{equation}
with $Y_{\ell m} (\theta,\varphi)$ being the spherical harmonics. With this decomposition, the scalar field perturbation equation (\ref{ScalarEq2}) becomes
\begin{align}
    \frac{\sqrt{g(r) \chi(r)}}{r} \frac{d}{dr} \left[\sqrt{g(r) \chi(r)} r^2 \frac{d}{dr} \left(\frac{\psi(r)}{r}\right)\right] + \left[\omega^2 + g(r) \left(-\frac{\ell(\ell+1)}{r^2} + \alpha {\cal G}\right)\right] \psi(r) = 0,
\end{align}
where the Gauss-Bonnet term evaluated on the background as
\begin{align}
    {\cal G} = \frac{2 g'(r) \left[(3 \chi(r) - 1) g(r) \chi'(r) - (\chi(r) - 1) \chi(r) g'(r)\right] + 4 (\chi(r) - 1) \chi(r) g(r) g''(r)}{r^2 g(r)^2}.
\end{align}
In the Schwarzschild limit, $g(r) = \chi(r) = 1 - 2 M/r$ and ${\cal G} = 48 M^2/r^6$.
By introducing the tortoise coordinate $dx = dr /\sqrt{g(r) \chi(r)} $ that maps the domain $r \in (r_0,\infty)$ to $x \in (-\infty,\infty)$, the above equation can be cast in a Schrodinger form. 
\begin{equation}
    \frac{d^2}{dx^2}\psi + \left[\omega^2 - V(r)\right]\psi = 0\label{SchrodingerEq},
\end{equation}
with the effective potential
\begin{align}
    V(r) = g(r) \left(\frac{\ell(\ell+1)}{r^2} - \alpha {\cal G}\right) + \frac{1}{2 r} \frac{d}{dr} \left(g(r) \chi(r)\right).
\end{align}

According to a well-known result in quantum mechanics, a sufficient condition for the existence of unstable modes of the Schrodinger-like equation (\ref{SchrodingerEq}) is \cite{Doneva:2017bvd}
\begin{equation}
{\cal I} \equiv  \int_{-\infty} ^{\infty} V(r) dx = \int_{r_0}^\infty \frac{V(r)}{\sqrt{g(r) \chi(r)}} dr < 0.
\label{condition}
\end{equation}
It should be emphasized that this condition is sufficient but not necessary, so there may still be unstable modes even if ${\cal I}>0$. However, this criterion can serve as a useful indicator to predict the onset of tachyonic instability. When the condition is only slightly violated, instability is likely to occur, whereas a significant violation will prevent the instability. In the parameter space, it can help us to carve out a region where the tachyonic instability is sufficient to be triggered. 

In the limit $\epsilon=0$, the metric is reduced to the standard Schwarzschild metric and in this case, there exists a threshold value $\alpha = 0.726$, above which the tachyonic instability will be triggered for the $\ell=0$ mode \cite{Zhang:2020pko}. To make a comparison with this known result and for simplicity, in the following, we will fix $\alpha = 0.726$ and $\ell=0$ unless specifically stated otherwise, and focus on the influences of the deformation parameters $\{\epsilon, a_2, b_2\}$. Also, we set $M = 1$ so that all quantities are measured in units of $M$. 

\begin{figure}[!htbp]
        \centering
        \includegraphics[width=0.6\textwidth]{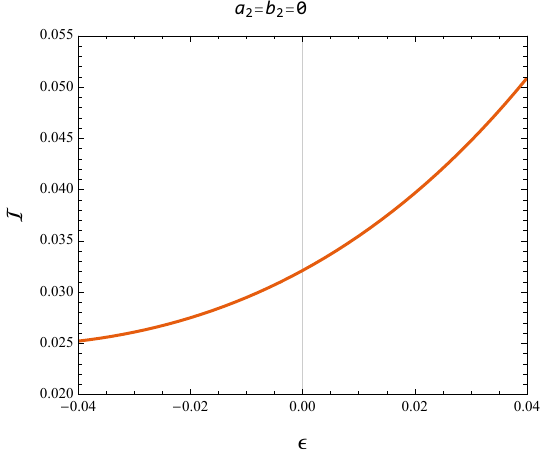}\quad
        \caption{${\cal I}$ as a function of the horizon deviation parameter $\epsilon$ with $a_2=b_2=0, \alpha = 0.726$ and $\ell=0$.}
        \label{fig:V1}
    \end{figure}

With the sufficient condition (\ref{condition}), we first analyze the effect caused by the horizon deviation parameter $\epsilon$ alone. In Fig. \ref{fig:V1}, the integral ${\cal I}$ in (\ref{condition}) is shown as a function of $\epsilon$ with $a_2=b_2=0$. It can be seen that in the range $\epsilon \in [-0.04,0.04]$, ${\cal I}$ increases slowly with increasing $\epsilon$ and is always positive, so the sufficient condition (\ref{condition}) cannot be satisfied. However, we can see that ${\cal I} (\epsilon < 0) < {\cal I} (\epsilon=0)$ which indicates that the instability may still be triggered for $\epsilon<0$ as confirmed by the numerical simulations below. 

\begin{figure}[!htbp]
        \centering
        \includegraphics[width=0.45\textwidth]{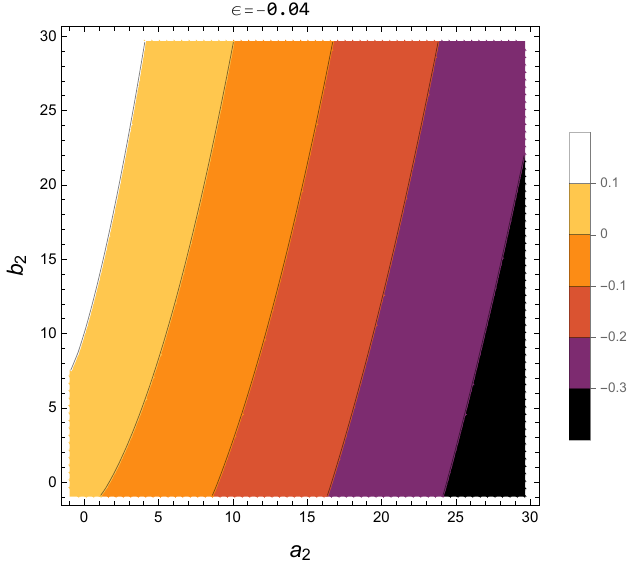}\quad
         \includegraphics[width=0.45\textwidth]{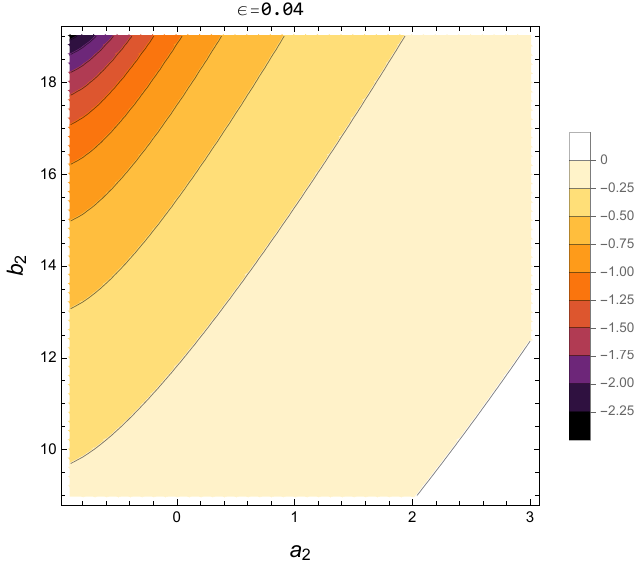}
        \caption{Density plot of ${\cal I}$ in the $a_2-b_2$ plane for $\epsilon = -0.04$ and $0.04$, respectively. We fix $\alpha=0.726$ and $\ell=0$.}
        \label{fig:V2}
\end{figure}

In Fig. \ref{fig:V2}, we plot the value of ${\cal I}$ in $a_2 - b_2$ plane for two typical values of $\epsilon$. In the left panel with $\epsilon<0$, it can be seen that the sufficient condition can be more easily satisfied by larger $a_2$ and smaller $b_2$. And, it should be noted that the sufficient condition cannot be satisfied for $a_2<1$. This indicates that larger $a_2$ or smaller $b_2$ will induce more violent instability. Furthermore, the effect of $a_2$ on ${\cal I}$ is much greater than that of $b_2$. However, the situation changes when $\epsilon>0$ as shown in the right panel, where one can see that the sufficient condition can be more easily satisfied by smaller $a_2$ and larger $b_2$.

\subsection{Time evolution of the scalar field perturbations}

Now let us study the scalar field perturbation in time domain. We will apply the numerical strategy as in Refs. \cite{Krivan:1996da,Pazos-Avalos:2004uyd,Dolan:2011dx,Doneva:2020nbb,Doneva:2020kfv,Zhang:2022sgt} to solve the scalar field perturbation equation (\ref{ScalarEq2}). By decomposing the scalar field, $\phi(t,x,\theta,\varphi)=\sum_{l,m}\Psi(t,x) Y_{lm}(\theta,\varphi)$, and introducing an auxiliary variable $\Pi \equiv \partial_t \Psi$, the scalar field perturbation equation can finally be cast in the following form
\begin{align}
    \Pi &= \partial_t \Psi,\nonumber\\
    \partial_t \Pi &= \partial_x^2 \Psi + \frac{2 \sqrt{g(r) \chi(r)}}{r} \partial_x\Psi + g(r) \left[-\frac{\ell(\ell+1)}{r^2} + \alpha {\cal G}\right]\Psi.
\end{align}
The form presented is well-suited for the method of line. Specifically, we utilize the fourth-order Runge-Kutta integrator for conducting simulations in the time direction, along with employing a finite difference scheme for the spatial derivative. Through adjusting the grid sizes, we have assessed the precision of the numerical simulations.

We consider the initial scalar field perturbation to be a Gaussian wave-packet localized outside the horizon at $x=x_c$ with width $\sigma$ and has time symmetry, 
\begin{align}
    &\Psi (t=0, x) \sim e^{-\frac{(x - x_c)^2}{2 \sigma^2}},\nonumber\\
    &\Pi (t=0, x) =  0. 
\end{align}
In the following, without loss of generality, the observer is located at $x=20$.

\begin{figure}[!htbp]
    \centering
    \includegraphics[width=0.6\textwidth]{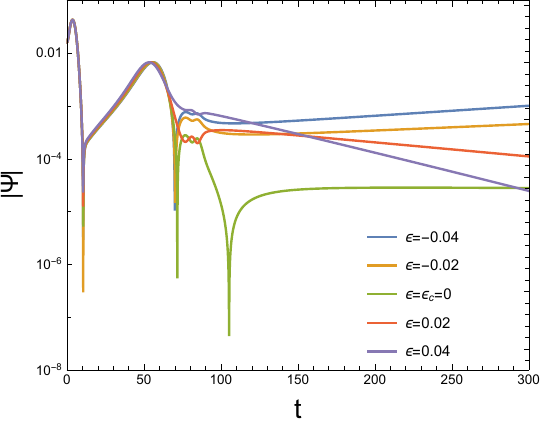}\quad
    \caption{(color online) Time evolutions of the scalar field perturbations for various values of $\epsilon$ with $a_2=b_2=0, \alpha=0.726$ and $\ell=0$. }
    \label{fig:Psi0}
\end{figure}

\begin{figure}[!htbp]
    \centering
    \includegraphics[width=0.45\textwidth]{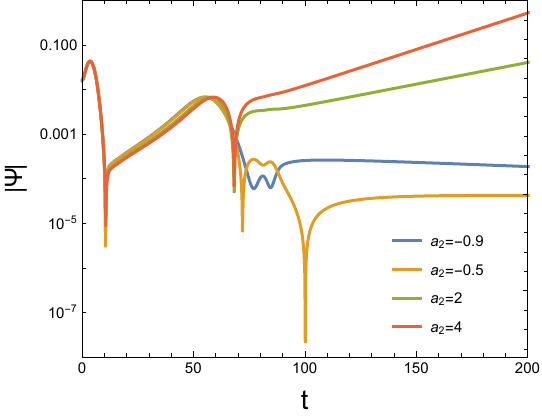}\quad
    \includegraphics[width=0.45\textwidth]{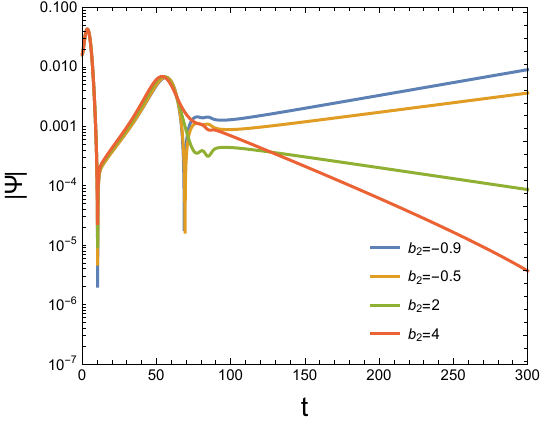}    
    \caption{(color online) Time evolutions of the scalar field perturbations for various values of $\{a_2, b_2\}$ with $\epsilon = -0.04, \alpha=0.726$ and $\ell=0$. In the left panel, we fix $b_2=0$, while in the right panel we fix $a_2=0$.}
    \label{fig:Psi1}
\end{figure}

\begin{figure}[!htbp]
    \centering
    \includegraphics[width=0.45\textwidth]{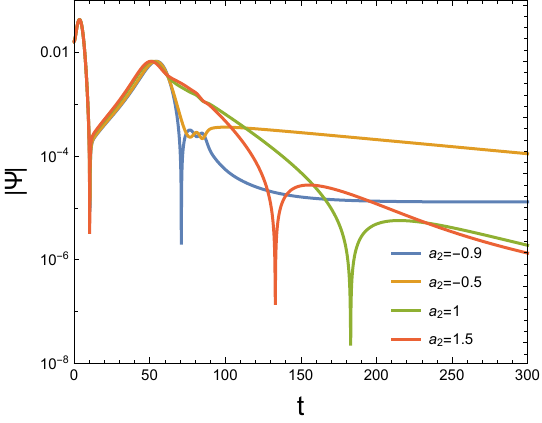}\quad
    \includegraphics[width=0.45\textwidth]{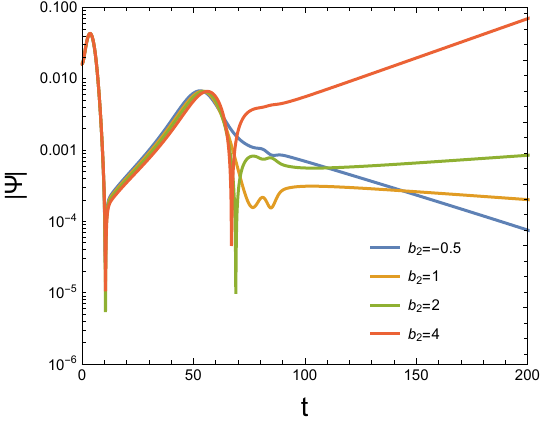}
    \caption{(color online) Time evolutions of the scalar field perturbations for various values of $\{a_2, b_2\}$ with $\epsilon = 0.04, \alpha=0.726$ and $\ell=0$. In the left panel, we fix $b_2=0$, while in the right panel $a_2=0$.}
    \label{fig:Psi2}
\end{figure}

The time evolutions of the scalar-field perturbations are shown in Figs. \ref{fig:Psi0}, \ref{fig:Psi1} and \ref{fig:Psi2} for typical values of the deformation parameters. In each figure, we fix two of the three deformation parameters and focus on the influence of the remaining parameter.

In Fig. \ref{fig:Psi0}, the influence of $\epsilon$ on the dynamics of the scalar field perturbation is shown. It can be seen that the tachyonic instability occurs (but moderately) when $\epsilon < \epsilon_c =0$, and becomes more violent as $\epsilon$ decreases. This result confirms our previous conclusion from Fig. \ref{fig:V1}.


In Fig. \ref{fig:Psi1}, we fix $\epsilon = -0.04$. From the left panel, one can see that tachyonic instability occurs as $a_2$ increases to exceed some threshold value. Furthermore, larger $a_2$ induces more violent instability. From the right panel, one can see that tachyonic instability occurs as $b_2$ decreases to below some threshold value. Furthermore, a smaller $b_2$ induces more violent instability. It should be noted that at the threshold ${\cal I}$ remains positive and the sufficient condition is not satisfied.  However, when $\epsilon>0$, $\{a_2, b_2\}$ show opposite influences on the occurrence of tachyonic instability, as can be seen in Fig. \ref{fig:Psi2}. In this case, smaller $a_2$ or larger $b_2$ induces more violent instability. These observations confirm our previous conclusions from Fig. \ref{fig:V2}.

\subsection{Static and spherical scalar clouds}
The tachyonic instability will result in spontaneous scalarization, under which scalar cloud will form around the black holes. For simplicity, we work in the ``probe limit" and neglect the backreaction of the scalar field on the background. Assuming that the scalar cloud formed is static and spherical, e.g, $\phi=\phi(r)$, the scalar field equation (\ref{ScalarEq1}) becomes
\begin{align}
    \phi''(r) + \left[\frac{2}{r} + \frac{g'(r)}{2 g(r) } + \frac{\chi'(r)}{2 \chi(r) } \right] \phi'(r) + \frac{\alpha}{\chi(r)} \frac{d f(\phi)}{d\phi}  {\cal G} = 0. \label{ScalarEq3}
\end{align}
With this equation, we can obtain the asymptotic solution of the scalar field near the event horizon $r=r_0$,
\begin{align}
    \phi(r) = \phi_0 - \frac{\alpha}{\chi'(r_0)} \frac{d f(\phi_0)}{d\phi} {\cal G}(r_0) (r-r_0) + {\cal O}((r-r_0)^2), \qquad r \rightarrow r_0, \label{SeriesAtHorizon}
\end{align}
where the constant $\phi_0$ is the value of the scalar field at the event horizon. The scalar field equation (\ref{ScalarEq3}) possesses a scaling symmetry under $\phi \rightarrow \lambda \phi$ and $\beta \rightarrow \lambda^{-2} \beta$, with which we can fix $\phi_0=1$. With (\ref{SeriesAtHorizon}), the scalar field equation (\ref{ScalarEq3}) can be solved from a point sufficiently close to the horizon towards infinity. The solution is determined by a set of five parameters $\{\alpha, \beta, \epsilon, a_2, b_2\}$. When the values of $\{\beta, \epsilon, a_2, b_2\}$ are fixed, only specific discrete values of $\alpha$ allow for a bounded solution where $\phi(r \rightarrow \infty) = 0$ to be achieved. Actually, for fixed $\{\beta, \epsilon, a_2, b_2\}$, there exists an infinite countable set of the coupling constant, $\{\alpha(\beta, \epsilon, a_2, b_2;n)\}_{n=0}^\infty$, which can support the bounded scalar cloud with $n$ labeling the number of nodes of the solution.

\begin{figure}[!htbp]
    \centering
    \includegraphics[width=0.7\textwidth]{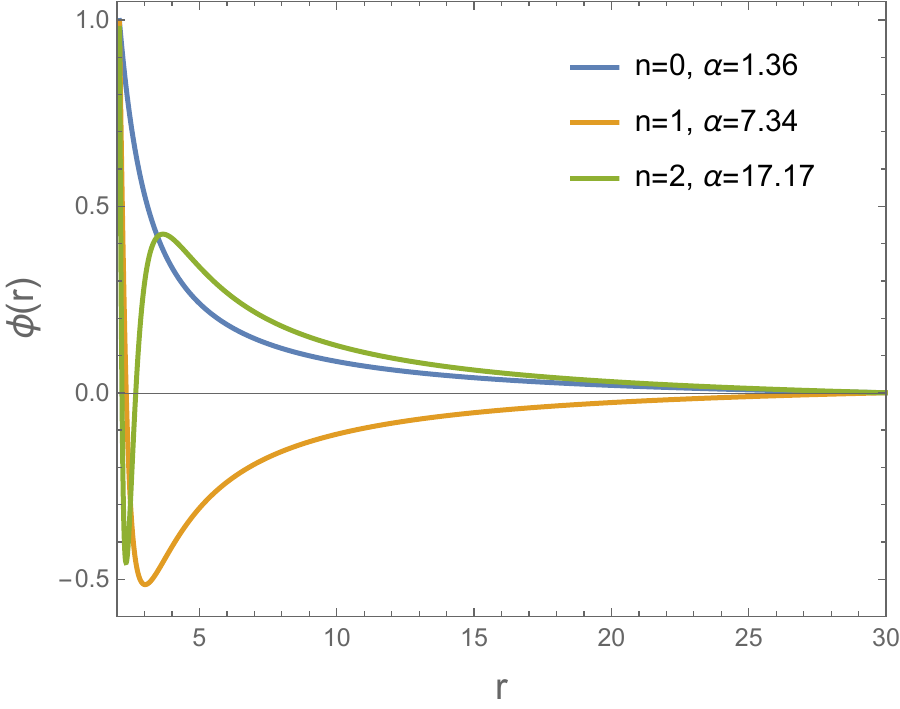}
    \caption{(color online) Configurations of the scalar clouds with zero, one and two nodes respectively (labeled by $n$). We set $\epsilon=-0.04$ and $a_2= b_2= \beta=1$. The value of $\alpha$ is determined correspondingly by the bounded-solution condition $\phi(r\rightarrow \infty)=0$.}
    \label{phirKZ1}
\end{figure}

In Fig. \ref{phirKZ1}, typical configurations of the scalar clouds with zero, one and two nodes are shown respectively. We set $\epsilon=-0.04$ and $a_2= b_2= \beta=1$. The value of $\alpha$ is determined correspondingly by the bounded-solution condition. It can be seen that the formation of the scalar cloud with more nodes requires larger $\alpha$.

\begin{figure}[!htbp]
    \centering
    \includegraphics[width=0.32\textwidth]{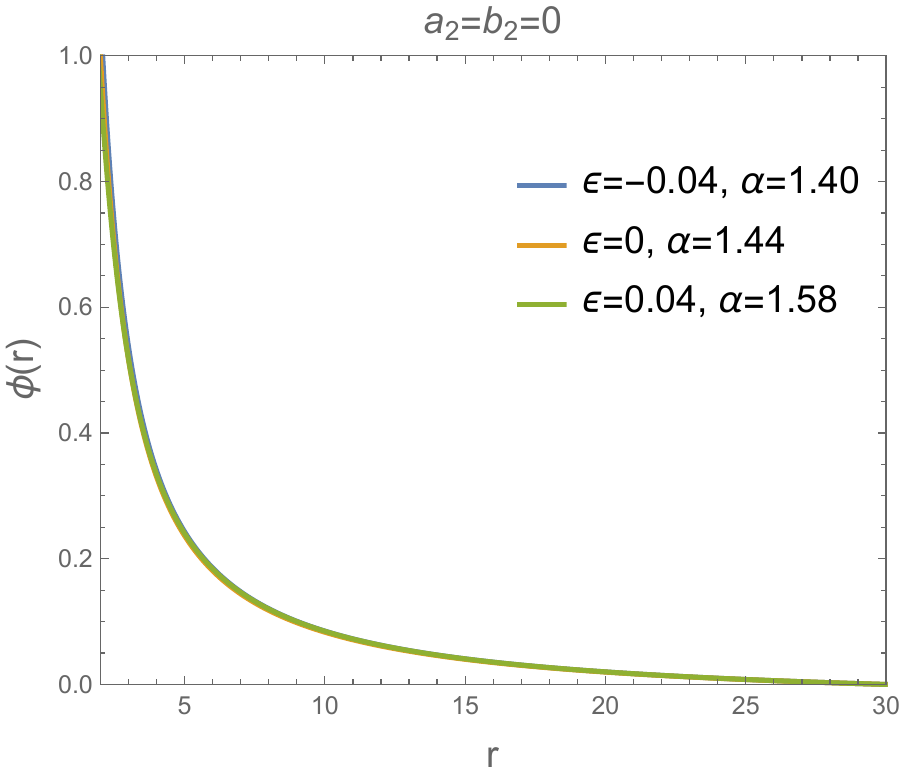}
    \includegraphics[width=0.32\textwidth]{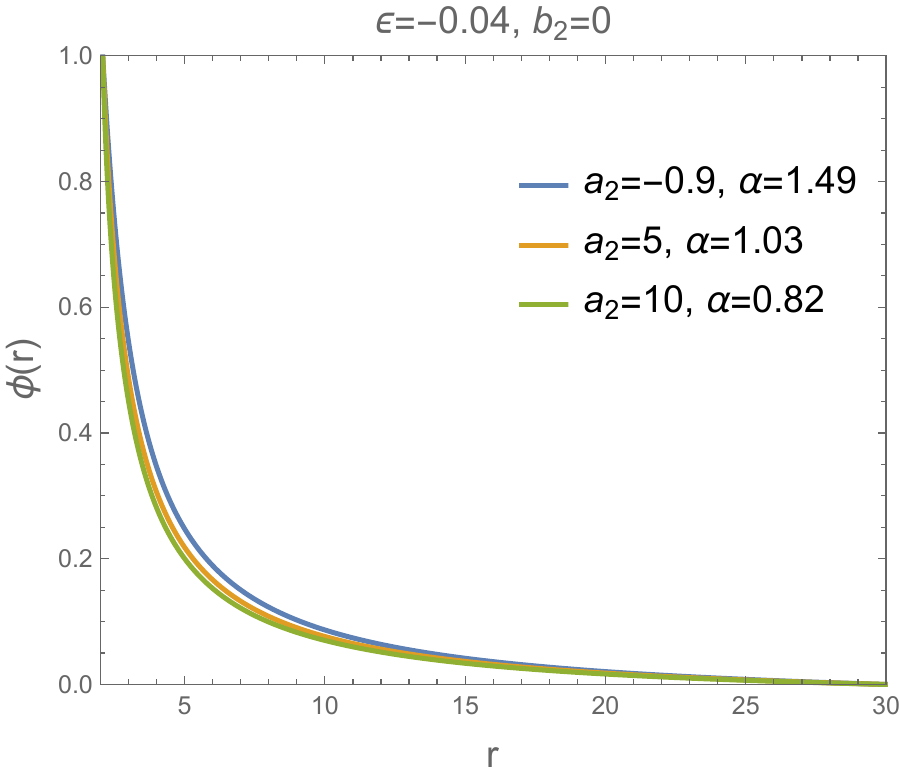}
    \includegraphics[width=0.32\textwidth]{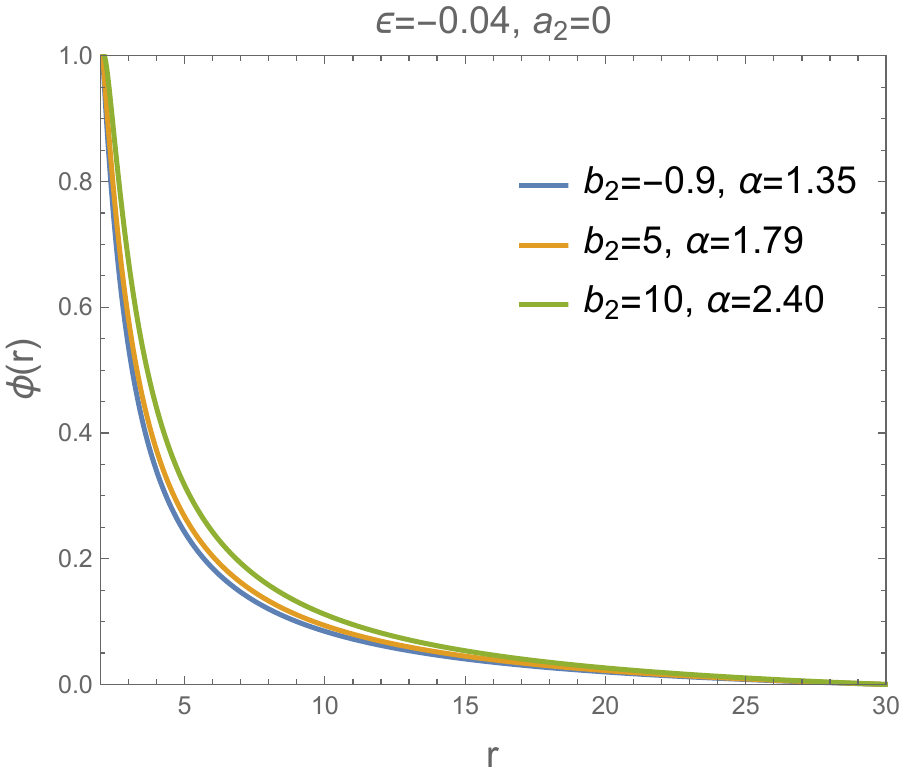}
    \caption{(color online) Configurations of the scalar clouds with zero nodes for various values of the deformation parameters. We set $\beta=1$. The value of $\alpha$ is determined correspondingly by the bounded-solution condition $\phi(r\rightarrow \infty)=0$. In the left panel, it is observed that the configurations of the scalar clouds with different $\epsilon$ exhibit significant overlap.}
    \label{phirKZ2}
\end{figure}

To show the influence of the deformation parameters on the scalar cloud, we focus on the scalar cloud with zero nodes. In Fig. \ref{phirKZ2}, configurations of the scalar cloud with zero nodes for various values of the deformation parameters are shown. From the figure, it can be seen that scalar clouds with different deformation parameters take similar configurations, with the primary distinction found in the necessary value of the coupling constant $\alpha$. It can be seen that smaller $\alpha$ is required for smaller $\epsilon$ when $a_2$ and $b_2$ are fixed. Furthermore, for a negative value of the horizon deviation parameter, i.e., $\epsilon=-0.04$, smaller $\alpha$ is required for larger $a_2$ or smaller $b_2$.  For a positive $\epsilon$, we have similar scalar cloud configurations, and it is found that smaller $\alpha$ is required for smaller $a_2$ or larger $b_2$. These results agree with previous conclusions.

\section{Johannsen-Psaltis metric}
Now, we are going to study the wave dynamics of scalar field perturbations on the background of another parameterized Schwarzschild-like metric, the Johannsen-Psaltis metric \cite{Johannsen:2011dh}
\begin{align}
    g (r)&=N(r) [1+h(r)],\\
    \chi (r)&=N(r) [1+h(r)]^{-1},\\
    N(r)&=1-\frac{2M}{r},\\
    h(r)&=\sum_{n=0}^{\infty}k_n \left(\frac{M}{r}\right)^n\label{h}.
\end{align}
As the Schwarzschild black hole, the event horizon remains at $r=2M$ without shift. However, the near-horzion geometry is deformed with an infinite number of
deformation parameters $\{k_n\}$. In the limit of all $k_n \rightarrow 0$, the metric recovers the Schwarzschild metric exactly.  

\begin{figure}[!htbp]
        \centering
        \includegraphics[width=0.6\textwidth]{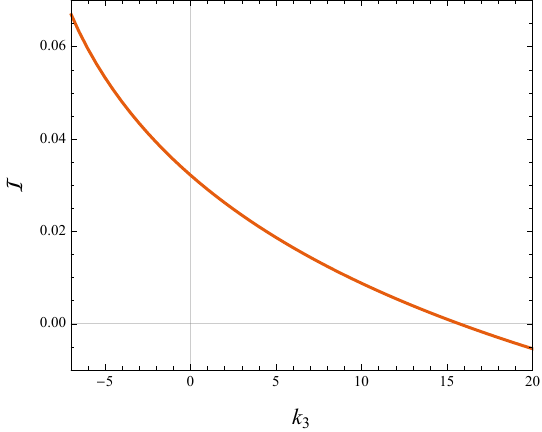}\quad
        \caption{${\cal I}$ as a function of $k_3$ with $\alpha = 0.726$ and $\ell=0$.}
        \label{fig:V3}
    \end{figure}

For this metric to serve as a good mimic of the Schwarzschild black hole in the weak-field region, we should have $k_0=k_1=0$. Moreover, the value of $k_2$ is strictly constrained by current astrophysical observations to be extremely close to zero, i.e.,  $|k_2| \leq 4.6\times 10^{-4}$ \cite{Williams:2004qba,Johannsen:2011dh}, therefore we neglect its effect and set $k_2=0$. Considering that the deformations from the high-order terms of $h(r)$ are small, we take $k_n=0$ for $n>3$ and focus on the effects caused by the leading deformation parameter $k_3$ alone. Then, the function $h(r)$ reduces to
\begin{equation}
    h(r)=k_3 \left(\frac{M}{r}\right)^3.
\end{equation}
Note that tests of general relativity with current astronomical observations place no constraints on the value of $k_3$ \cite{Psaltis:2008bb}. However, to make sure that the metric functions are always positive outside the event horizon and avoid the naked singularities, the parameter $k_3$ should satisfy
\begin{equation}
    k_3 > -8.\label{constraint1}
\end{equation}
This constraint ensures that the Gauss-Bonnet term always converges outside the event horizon. 

Similarly, with the condition (\ref{condition}), we can find out the range of parameters $k_3$ that fully guarantees the triggering of tachyonic instability. In Fig. \ref{fig:V3}, the integral ${\cal I}$ as a function of $k_3$ is plotted with $\alpha=0.726$ and $\ell=0$. It can be seen that ${\cal I}$ decreases with the increase of $k_3$, and ${\cal I}<0$ for $k_3 \gtrsim 15$. This means that larger $k_3$ may induce stronger instability. We should emphasize once again that this condition is sufficient but not necessary, so there may still be tachyonic instability for $k_3<15$. Actually, as ${\cal I}(k_3>0) < {\cal I}(k_3=0)$, the tachyonic instability will be triggered for $k_3>0$ as confirmed by the numerical simulations below.

\begin{figure}[!htbp]
        \centering
        \includegraphics[width=0.7\textwidth]{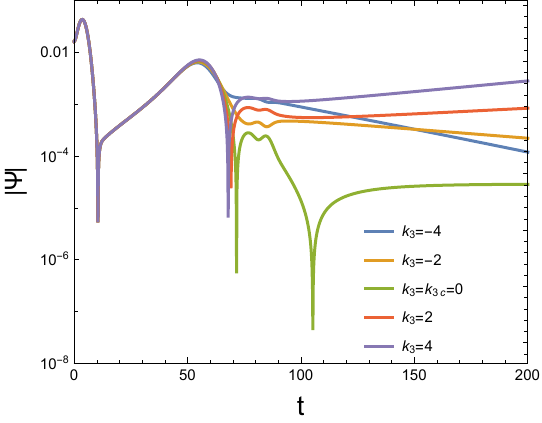}\quad
        \caption{Time evolutions of the scalar field perturbations for various values of $k_3$ with $\alpha=0.726$ and $\ell=0$.}
        \label{fig:Psi3}
\end{figure}

\begin{figure}[!htbp]
    \centering
    \includegraphics[width=0.47\textwidth]{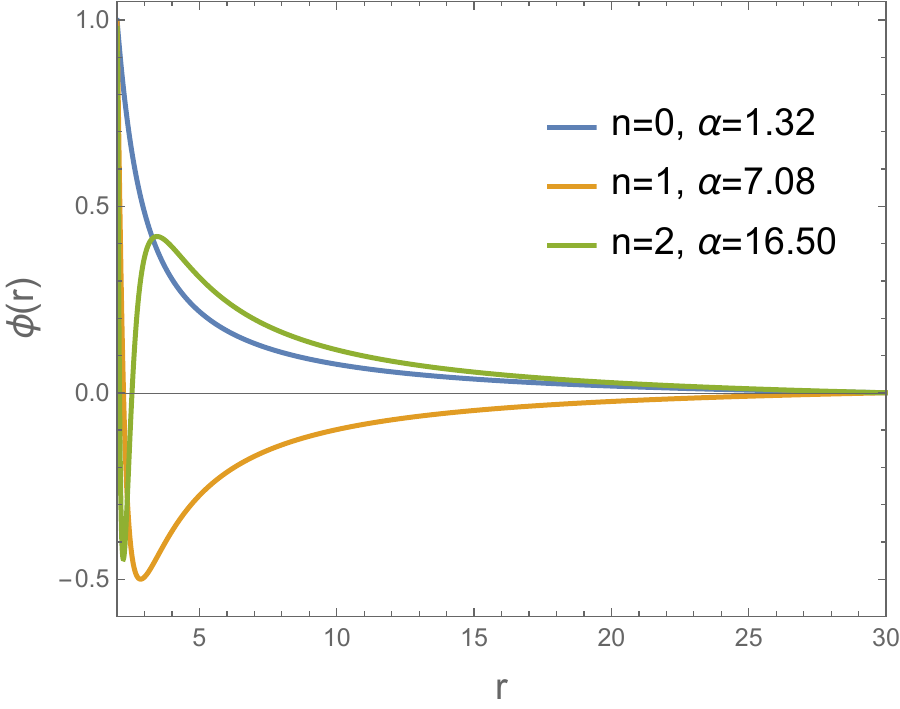}\quad
    \includegraphics[width=0.47\textwidth]{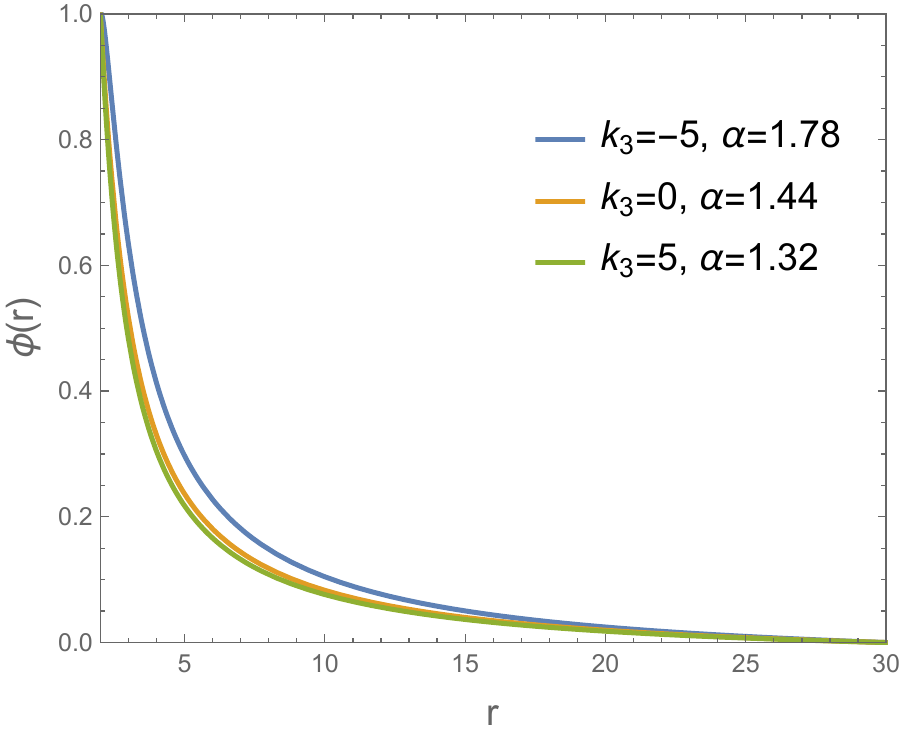}
    \caption{(color online) Configurations of the scalar clouds with $\beta=1$. In the left panel, we set $k_3=5$ and show the scalar clouds with zero, one and two nodes respectively. In the right panel, we focus on the scalar clouds with zero nodes. }
    \label{phirJP}
\end{figure}

Following the same numerical strategy as in the KZ case, the time evolution of the scalar field perturbations on the JP background can also be studied, as shown in Fig. \ref{fig:Psi3}. It can be seen that tachyonic instability occurs when $k_3$ exceeds a threshold value $k_{3c}$. Note that $k_{3c} \simeq 0$. It can be understood that when $k_3=0$, the JP metric reduces to the Schwarzschild black hole exactly, and in this case the perturbation will reach a critical state that has been mentioned above. Furthermore, a larger $k_3$ induces more violent instability. This result agrees with observations from Fig. \ref{fig:V3}. 

Samples of the scalar clouds resulting from the tachyonic instability are shown in Fig. \ref{phirJP}. From the right panel, it can be seen that smaller $\alpha$ is required for larger $k_3$. This confirms our previous conclusion that tachyonic instability and spontaneous scalarization favor larger $k_3$.

\section{Summary and Discussions}

This study investigates the occurrence of spontaneous scalarization in parameterized Schwarzschild-like black holes. Two parametrized metrics, namely the KZ and JP metrics, are examined. These metrics can effectively resemble the standard Schwarzschild black hole in the weak-field regime and are consistent with current astronomical observations. Their distinguishing feature from the standard Schwarzschild black hole lies in their near-horizon structures. Therefore, the phenomenon of spontaneous scalarization, which is closely related to the near-horizon structure of spacetime, is anticipated to differentiate between them. 

Both metrics contain several deformation parameters introduced to characterize the deviations from the standard Schwarzschild black hole especially in the near-horizon region. By examining the integral of the effective potential (\ref{condition}), and performing the time evolution of the scalar field perturbation, it is found that the deviations have a significant influence on the occurrence of the tachyonic instability. 

In the KZ metric, the occurrence of tachyonic instability favors a more negative $\epsilon$. The explicit influence of the other two deformation parameters, $\{a_2, b_2\}$, depends on the sign of $\epsilon$. When $\epsilon<0$, the tachyonic instability occurs easier for larger $a_2$ or smaller $b_2$; While $\epsilon>0$, they have the opposite effect.

The JP metric is more concise in form and contains only one parameter, $k_3$, in the leading order. It is found that the tachyonic instability occurs more easily for larger $k_3$.

The tachyonic instability results in spontaneous scalarization under which a scalar field accumulates around the black holes. By solving the scalar field equation in the probe limit, we show the existence of scalar clouds that are finally formed. For different values of the deformation parameters, the scalar clouds have similar profiles. However, it is important to note that the value of the coupling constant $\alpha$ required for their formation is quite different. This means that for fixed $\alpha$,  whether a scalar cloud can form depends on the explicit values of the deformation parameters. This suggests a possible way to test the parameterized black holes and thus the Kerr hypothesis by observing the phenomenon of spontaneous scalarization.

\begin{acknowledgments}

	This work is supported by the National Natural Science Foundation of China (NNSFC) under Grant No 12075207.

\end{acknowledgments}

\bibliographystyle{utphys}
\bibliography{Refslib}
\end{document}